# Towards better representation of Indian summer monsoon rainfall in CMIP6 models: Evaluation of MISO and MJO simulation

Ushnanshu Dutta[1*], Moumita Bhowmik[2], Anupam Hazra[1,2*], Chein-Jung Shiu[3] and Jen-Ping Chen[1]

[1] Department of Atmospheric Sciences, National Taiwan University, Taipei, Taiwan

[2] Indian Institute of Tropical Meteorology, Ministry of Earth Sciences, India

[3] Research Center for Environment Change, Academia Sinica, Taiwan

**Corresponding author(s):**

Dr. Ushnanshu Dutta (duttaushnanshu@ntu.edu.tw)

Dr. Anupam Hazra (hazra@tropmet.res.in)

**Key Points:**

- Fidelity test of four CMIP6 models are performed to simulate Indian summer monsoon oscillation and MJO.

- TaiESM1 performs better among the models in general. IITM-ESM also performs good in some aspects.

- Results highlight the importance of proper representation of cloud and convection in CMIP6 models.

# Abstract

The seasonal prediction of the Indian summer monsoon (ISM) and Monsoon Intraseasonal Oscillations (MISO), as well as the Madden Julian Oscillations (MJO) that strongly modulate MISO, is important to the country for water and crop management. We have analyzed the precipitation, convection, and total cloud fraction (TCF) in the sixth Coupled Model Intercomparison Projects (CMIP6). This study highlights the significant differences in simulating MISO and MJO between the two groups of selected CMIP6 models and physical reasons behind them. The mean and intraseasonal features of MISO and MJO varied significantly in CMIP6 models, which are linked with a better depiction of convection and total cloud fraction. The probability distributions of rainfall and OLR in CMIP6 models indicate significant variations in simulating ISM precipitating clouds. TaiESM1 and IITM-ESM demonstrate improvements in capturing the MISO features. TaiESM1 depicts better eastward propagation of MJO during both summer and winter. The biases in OLR and TCF are also less in the IITM-ESM and TaiESM1 models than in CanESM5 and FGOALS-g3. The results demonstrate the importance of cloud and convection in CMIP6 models to depict realistic MISO and MJO and provide a road map for improving ISM climate prediction and projections.

## 1. Introduction

The prediction and projection of seasonal mean (June-September) Indian summer monsoon rainfall (ISMR) have strong impact for a variety of socio-economic reasons. Over hundred years, there has been attempted ISMR prediction, essential for farmers, financial markets, and policy makers alike, firstly with empirical techniques, and more recently with coupled climate models along with earth system models for the accurate long lead forecast. But there is limited success so far after continuous efforts. Why has ISMR remained a challenging climate system to predict? What present generation CMIP6 models have difficulty in simulating monsoon? Whether CMIP6 models can depict MISO and MJO, related to ISMR simulation? The seasonal prediction of ISM and Monsoon Intraseasonal Oscillations (MISO) are of much importance to the agriculture sectors. Madden Julian Oscillations (MJO) (Madden and Julian, 1972) also strongly modulates monsoon ISOs (MISO) (Pai et al., 2011, Joseph et al., 2009, Yasunari 1980, Hendon and Liebmann 1990, Straub et al., 2006, etc.). It is known that there is robust relationship between northward propagation of ISO and MJO (Wu et al. 1999). The evolution of ENSO can also be modulated by the eastward propagation of MJO convective activities (Lau and Shen, 1988; Weickmann, 1991), which again controls sub-seasonal oscillations (i.e., MISOs and synoptic activity) as demonstrated by Saha et al., (2019). It is therefore, crucial to better simulate these two tropical oscillations (ISO and MJO) in Coupled Global Climate Models (CGCM) to enhance the prediction skill of dynamical models (Rajeevan et al.,2012, Goswami and Goswami 2017, Jiang et al., 2015, Yang and Wang 2019).

The seasonal movement of the monsoon rain band known as the Inter-tropical Convergence Zone (ITCZ) is important for the ISM climate to modulate the wet summer and dry winter over the Indian continent (Goswami et al., 2003). Wang and Xu, (1997) have shown that prominent sub-seasonal fluctuations in the climatological rain-band and the circulation, which dynamically interacts with the SSTs is coupled with the seasonal rainfall over Indian subcontinent. The year-to-year and sub-seasonal variability of the seasonal and daily mean rainfall respectively over India are two most important variations in South Asian monsoon rainfall. During the 'active' ('break') monsoon period there is a prolonged above (below) sub-seasonal activity over Indian subcontinent and the seasonal mean may be related to the tendency and intensity of these spells. Therefore, MISOs can be treated as building blocks of the seasonal mean monsoon climate (Goswami, 2012) as the seasonal mean and its variability also characterized by the temporal clustering of synoptic events (Goswami et al., 2003).

The world global modeling community continuously improving the model physics in the parameterization schemes for the better representation of monsoon, which is evolved in many phases of CMIP models and CMIP6, is the latest model in this regard. The CMIP6 is expected to deliver the model's ability to characterize the monsoon and its variability, how clouds have impact on monsoon dynamics (Bony et al. 2015; Eyring et al. 2016), and how the future climate has changed (Gusain et al. 2020). The advanced model resolution and physics in CMIP6 models have shown advancement as reported by many previous studies (Sperber et al. 2013; Seo et al. 2013; Eyring et al. 2016; Xin et al. 2020). Dutta et al. (2022a) have recently demonstrated that clouds over south Asian monsoon region and their teleconnection with global predictors are better simulated in CMIP6 as compared to CMIP5. It is important to note that there are also differences among various CMIP6 models in terms of their performance in depicting mean monsoon. The better representation of convective to large-scale (stratiform) rain is important for the improved simulation of seasonal mean monsoon rainfall over India (Pokhrel and Sikka 2013, Hazra et al. 2017a, 2017b, 2020; Dutta et al. 2021). Recently, Kejun et al., (2021) have shown that greenhouse gases is one of the important factors in increasing the surface warming, where clouds and water vapor are interrelated to each other. Therefore, it will be judicious to make comparisons between two groups of CMIP6 models, (i) relatively better mean monsoon characteristics, (ii) underestimation of mean ISMR. This study will help for the better understanding of CMIP6 models, which will help to diagnose model's capability in terms of rainfall simulations (component wise convective and stratiform) over East Asian Monsoon Region.

There is rising trend in the extreme rainfall events over the Indian subcontinent under global warming scenario (Roxy et al. 2017). In recent years, several parts of India have experienced devastating extreme rainfall events, such as heavy rainfall over Mumbai on 26 July 2005 (Prasad and Singh 2005), Uttarakhand on 17 June 2013 (Ranalkar et al., 2016; Hazra et al., 2017c), Kerala on August 2018 (Hunt and Menon, 2020) that lead to unparalleled damage to life and property (Dube et al. 2014; Martha et al. 2015). Furthermore, several researchers argued that while the overall intensity and frequency of extreme events are increasing over the region with increasing spatial variability (Goswami et al. 2006; Roxy et al., 2017) and the further intensification of extreme precipitation over most parts of the subcontinent. Therefore, the need for a better understanding of cloud fraction and rainfall, which is related to the northward and eastward propagation of MISO and MJO respectively, motivated the present study. The relationship between the intraseasonal fluctuations (active and break spells) of ISMR with the phase propagation and amplitude of MISO and MJO are analyzed using results from CMIP6 models. Our objective is to find the physical process in general and cloud properties in particular to depict the mean state ISMR as well as the variability of these two tropical oscillations.

The later sections are organized as follows. Section 2 describes Data and Methodology. Section 3 contains the results of models and their evaluation with observation/reanalysis data. We conclude the study in Section 4.

**2. Data and Methodology:**

Rainfall, outgoing longwave radiation (OLR), total cloud fraction (TCF), specific humidity (SH), and wind data are obtained from four CMIP6 (ensemble id: r1i1p1f1) models, i.e., CanESM5 (Swart et al., 2019), FGOALS-g3 (Li et al., 2020), IITM-ESM (Swapna et al., 2015) and TaiESM1 (Lee et al., 2020) from the historical simulations of available last thirty years (1985-2014). Validations of model data sets are performed with observation/reanalysis data, with rainfall from the Global Precipitation Climatology Project Version 2.3 (GPCP, Adler et al., 2003), whereas TCF, SH, and wind data are taken from the recently released fifth generation of the European Centre for Medium-Range Weather Forecasts (ECMWF) reanalysis (ERA5; Hersbach et al., 2020). Dutta et al. (2022a) have documented the fidelity of TCF data from ERA5 with the observational data. Therefore, it is useful to use ERA5 data for long-term analysis due to observational data shortage. For OLR, the widely used National Oceanic and Atmospheric Administration (NOAA) interpolated data (Liebmann and Smith, 1996) are considered. Observational/Reanalysis data sets for climatological analysis are used for the year 1985-2014. However, GPCP daily data for rainfall are available only from 1997. Hence, the last 25 years of daily data (1997-2021) are used for the evaluation of intraseasonal analysis.

The primary goal of this study is to evaluate the performance of several CMIP6 models in depicting two tropical oscillations (i.e., MISO and MJO), which are important for seasonal mean ISMR. Four CMIP6 models are selected to compare the performance in simulating MISO and MJO based on Dutta et al., (2022a) study. The models IITM-ESM and TaiESM1 have relatively low biases in rainfall and total cloud fraction (TCF) as compared to CanESM5 and FGOALS-g3. Both CanESM5, and FGOALS-g3 models (hereafter called Group-I) have similar rainfall biases (dry bias > 6 mm/day) with strong underestimation of TCF more than 25%, (Fig. 1). On the other hand, less dry bias is seen for IITM-ESM (~ 3 mm/day) and TaiESM1 (~1 mm/day). TCF bias is also about 5% (Fig. 1) for these two models (hereafter called Group-II).

Now, the question is: whether the CMIP6 models having better seasonal mean (rainfall and TCF) also provide improved tropical oscillations (i.e., MISO and MJO)? To understand the behavior of the Indian summer monsoon (ISM) quantitatively, we analyzed the mean spatial pattern of rainfall, TCF, outgoing longwave radiation, northward and eastward propagation of rainfall and OLR and pattern correlation (map-to-map correlation) between the models and observational dataset. The probability distribution of rainfall and the simulation of convective to total rainfall ratio are also assessed. Active break spell performance of the models is also analyzed with the method following Dutta et al. (2020), i.e., using average daily rainfall anomaly over the central Indian core monsoon region (i.e., averaged over 74° E–86° E and 16° N–26° N) divided by the standard deviation. The active and break spells are considered as the periods during which the standardized daily rainfall (unfiltered) anomaly is more than $+1.0$ and less than $-1.0$, respectively, for three or more consecutive days (Rajeevan et al. 2010). Intraseasonal oscillations (ISOs) in different bands were calculated by applying the Lanczos band-pass filter (20–100 days, 10–20 days, and 30–60 days) over daily rainfall anomaly.

## 3. Results and Discussions:

### 3.1. Monsoon Analysis

#### 3.1.1 Mean State of Rainfall, Convection, and Clouds:

The June to September (JJAS) climatology of rainfall over global tropics from GPCP is shown in Figure 2a, whereas the same from the four models are shown in Fig. S1. To understand how the models can capture the mean pattern over the global tropics, we compared their pattern correlation (PC) with the GPCP data (Fig. 2a). All the models have strong PC (Fig. S1) with the observation, with PC coefficients (PCC) in the range of 0.75 to 0.86. The mean value of rainfall over the global tropics (0˚-360˚, 30˚S-30˚N) from observation is ~3 mm/day which is slightly overestimated by all the models. The model results differed significantly (Fig. S2a) over the key tropical locations defined by Goswami and Goswami (2017) e.g., Extended Indian Monsoon Region or EIMR (70˚E-100˚E, 10˚N-30˚N), West Pacific (140˚E-160˚E, 0-10˚N), Amazon and its surroundings (75˚W-45˚W, 15˚S-10˚N) (Fig. S2a). Over the EIMR, the TaiESM1 is the closest to the observation. The other three models underestimated the mean value, where IITM-ESM being better among them. For the West Pacific, FGOALS-g3 overestimated and CanESM5 underestimated the mean rainfall, whereas IITM-ESM and TaiESM1 are close to the observation (Fig. S1, S2a). TaiESM1 also performs better in depiction of rainfall over the Amazon and its surroundings regions (Fig. S1, S2a).

A generic problem of earlier global models is the dry (wet) bias over land (ocean) for the monsoon rainfall over the Indian subcontinent and the Asian summer monsoon region (Rajeevan and Nanjundiah, 2009). The new generation models are also not an exception in this regard (Dutta et al., 2022a, Gusain et al., 2019). However, the amount of bias and spatial extent differed among the models. The rainfall bias of four models over the South Asian monsoon region (SAMR) with respect to observation (GPCP) is shown in Figure 3. It is evident that, over the major parts of Indian landmass region, all the four models are showing dry biases. However, the intensity and spatial extent has been visibly well reduced in IITM-ESM (Fig. 3c) and TaiESM1 (Fig. 3d) compared to CanESM5 (Fig. 3a) and Fgoals-g3 (Fig. 3b). For quantitative estimates, we have computed the rainfall bias over central India (74°E-86°E; 16°N-26°N) from the four models. Central India is chosen because of its topographical homogeneity and an important marker for monsoon statistics (Dutta et al., 2022a). Though all the four models show dry biases the magnitude significantly varied among the models (Table 1). CanESM5 (TaiESM1) shows the highest (least) dry bias. IITM-ESM has also shown improvement in reducing the dry bias compared to CanESM5 and Fgoals-g3 (Fig.1, Table 1). A similar estimate over the equatorial Indian Ocean (65°E-95°E,5°S-5°N) shows wet bias from CanESM5 and Fgoals-g3. Contrastingly, IITM-ESM and TaiESM1 show dry bias (Table 1) with less magnitude. The Bay of Bengal (BoB) is a deep convective region where many low-pressure systems originate and bring rain to Indian landmass (Goswami 1987).. The average bias of BoB convective heating region (82°E-95°E,5°N-20°N) showed dry bias in CanESM5 and Fgoals-g3, while IITM-ESM and TaiESM1 showed wet bias (Table 1). However, the magnitude of bias is less in IITM-ESM and TaiESM1 than that of CanESM5 and FGoals-g3. Hence, we can comment that the bias magnitude is less in ITTM-ESM and TaiESM1 for all regions, with TaiESM1 performed better in general. Also, the wet bias (Fig. 3d) over the oceanic regions surrounding the maritime continents and the rain density distribution over the central India (CI) region (Fig. 3e) is also improved in TaiESM1. Earlier studies have found that models have the tendency to overestimate the light rain and underestimate the heavier rain (Dutta et al., 2021). Among the four analyzed models, CanESM5 overestimates the light rain the most, while IITM-ESM performed the best followed by TaiESM1. On the heavy rainfall side, TaiESM1 can produce higher frequencies that better match the observation (Fig. 3e).

The convective activities are essential for monsoon circulations (Prasad and Verma, 1985). The outgoing longwave radiation (OLR) is a proxy of convection, with low (high) values signifying deep (shallow) convection (Murakami, 1980). The OLR minima are observed over the BoB (Figure 2b), which signifies the frequent occurrence of deep convection events over this region (Rajeevan, 2001). These convective activities over BoB during the monsoon work as a heat source (Joseph and Sijikumar, 2004). Therefore, to understand how these models perform in capturing the convection, we compared the OLR values of the models (Fig. S3a-d). The observation (NOAA) shows low OLR values over Arabian Sea (AS) and central to northeastern states of India (Fig. 2a), which is consistent with rainfall distribution (Fig. 2a). All the models can capture the convection zone over the BoB (Fig. S3a-d). However, over the Indian landmass, CanESM5 and FGOALS-g3 overestimated the OLR values (Fig. 4a, b), while they are better reproduced in IITM-ESM and TaiESM1 (Fig. 4c, d).

The quantitative comparison of OLR values of the models with observation over different parts of the ISM region (i.e., EIMR, BoB, CI and AS) is shown on Figure S2b. From the observation, we can find that the OLR value over the BoB is also the least among all the four regions. CanESM5 overestimated the OLR for all the regions. All four models slightly overestimated the OLR over BoB. IITM-ESM and TaiESM1 are closer to the observation for the rest of the selected regions i.e., EIMR, CI and AS. Over the global tropics, all models performed well in the mean spatial pattern of the OLR (Fig. 2b), as revealed by their respective PCC (Fig. S3a-d), with TaiESM1 being closest to the observation (PCC ~ 0.92). A quantitative estimate of the mean bias shows that all the models overestimated the OLR over the global tropics. The TaiESM1 shows the highest mean bias (~11.63 Watt/m$^2$) over the global tropics, and IITM-ESM shows the least bias (2.51 Watt/m$^2$).

The probability distribution of OLR over the greater ISM region (i.e., the EIMR) are plotted in Figure 4e. The OLR in CanESM5 and FGoals-g3 models overestimate the tail (highest OLRs) and highly underestimate the crest (lowest OLRs) (Fig. 4e). On the other hand, OLR PDFs in IITM-ESM and TaiESM1 are relatively better than the other two CMIP6 models (Figure 4e), but underestimation still exist in all CMIP6 models as compared to NOAA (Fig. 4e). It is to be noted that the higher (lower) OLR signify the deep (shallow) convection (Krishnamurti et al., 1989; Murakami, 1980). Therefore, too much (little) shallow (deep) convection in CanESM5 and FGoals-g3 models (Fig. 4e) is responsible for the ample to lighter rain (Fig. 3e), which is a generic problem of all latest generation CMIP6 model also. The PDFs of OLR is better in IITM-ESM and TaiESM1 models (Fig. 4e), which is also related to their better rainfall PDFs (Fig. 3e). In this regard, the physically-based microphysical parameterization using in situ observation can shed light on obtaining realistic probability distribution of OLR.

The distribution of clouds is also closely linked with the convection (Chaudhari et al., 2016) and plays key role for ISM not only in term of precipitation generation but also radiation influences. Gusain et al., (2019) showed that the TCF has been improved from CMIP5 to CMIP6 models. Our earlier study (Dutta et al., 2022a) showed that representation of the TCF over the SAMR has been improved in CMIP6 multi-model mean along with the teleconnection of TCF over ISM region with sea surface temperature over the different parts of the globe. In this study, we compare the spatial distribution of the TCF from these four models during the ISM. Figure 5. Shows that CanESM5 and FGOALS-g3 largely underestimates the TCF over the SAMR, while IITM-ESM and TaiESM1 showed substantial improvements. Over the Indian landmass, the bias is the least in TaiESM1 among other models (Fig. 5d). Quantitative estimate of the bias over EIMR also shows that all the models show negative bias. However, the magnitude of the bias is the least in TaiESM1 (-1.55%). The bias is also quite less in IITM-ESM (-7.2%) than the other two models (CanESM5: -19.20%; FGOALS-g3: -17.50%). The mean pattern of the TCF over the global tropics from the four models is shown in the supplementary (Fig. S3e-h). The pattern correlation of the four models with the reanalysis (Fig. 2c) are similar (0.7 ~ 0.8). Among the four models, CanESM5 and TaiESM5 are found to overestimate the cloud fraction in majority parts of global Tropical Ocean. Quantitative bias estimates over the global tropics revealed the same thing. CanESM5 (3.35%) and TaiESM1 (4.87%) showed a positive bias whereas FGOALS-g3 (-7.22%) and IITM-ESM (-0.95%) showed negative biases. However, the magnitude is the least in IITM-ESM.

### 3.1.2. Monsoon Intraseasonal Oscillation (MISO)

#### 3.1.2.1. Active -Break Spell:

The seasonal mean monsoon is largely determined by the active and break spells of monsoon (Rajeevan, et al., 2010). These active and break spells are the building blocks of MISO (Goswami and Ajayamohan, 2001). Sikka and Gadgil (1980) found that the active–break spells are largely caused by the northward-propagating, 30–60-day MISOs from the equatorial Indian Ocean to the ISM region. However, westward propagating fluctuations on time scale of 10–20 days or quasi bi-weekly mode (QBM) also contributes to the intraseasonal oscillations (Annamalai and Slingo 2001; Goswami 2007, Hazra et al., 2020). Here, we first analyze models' performance on active-break spell. During the active (break) spells, the majority of Indian landmass and BoB receive anomalously more (less) rainfall (Dutta et al., 2020, 2022b). The differences of rainfall between active and break spells from observation and the four models are shown in Figure 6. Positive (negative) differences over a region signify heavier rainfall during active (break) spell. Observation shows heavier rainfall over the Indian landmass and adjacent BoB, Arabian Sea during active spell and over the Equatorial Indian Ocean (EIO) during break spell (Fig. 6a). Though all the four models (Fig. 6b-e) can capture the basic feature of the active-break rainfall difference, they vary in spatial extent. CanESM5 (Fig. 6b) and FGOALS-g3 (Fig. 6c) can capture the rainfall difference over some parts of the Indian landmass and EIO with limited spatial extent. This spatial distribution of rainfall over the Indian subcontinent is improved in IITM-ESM (Fig. 6c) and TaiESM1 (Fig. 6d). The quantitative difference of observed rainfall between active and break spell over the Indian summer monsoon (ISM) region (70°E-90°E, 10°N-30°N) is around 10.5 mm/day. The FGOALS-g3 (~5 mm/day) and CanESM5 (~7 mm/day) both underestimates it largely, while better performances are achieved from IITM-ESM (~8.2 mm/day) and TaiESM1 (~10 mm/day). The OLR difference between active and break spell also shows that the TaiESM1 can capture the shift of convection band between the Indian landmass and EIO (Fig. S4e) as reflected in the difference in cloudiness (Fig. 7). The difference of TCF between active and break spell is also better captured in IITM-ESM (Fig. 7c) and TaiESM1 (Fig. 7d). The primary characteristic of the Indian monsoon circulation is marked by a robust low-level southwesterly jet known as the Findlater jet (Wilson et al., 2018). This jet reaches its peak strength in the vicinity of the Somali coast and the Arabian Sea region (Thompson et al., 2008, Jospeh and Sijikumar, 2004). The Southwesterly Jet (SWJ) serves as a connecting link between the Mascarene high and the Indian monsoon trough, constituting the lower component of the monsoon Hadley cell. The SWJ core exhibits greater strength during active wind composites and weaker strength during break wind composites at 850 hPa (Dutta et al., 2020). The difference of 850 hPa wind speed between active and break spell is also more realistic for the IITM-ESM and TaiESM1 among the four models (Fig. S5). The SWJ significantly impact the availability of moisture (Viswanadhapalli et al. 2020), which is further related to the intraseasonal variability of tropospheric humidity (Turner and Slingo, 2008). Height-latitude profile of specific humidity difference (averaged over 70°E-90°E) between active and break spell is shown in Figure 8. All the four models can simulate a higher humidity over the Indian landmass latitudes (Fig. 8a-e) during active spell as revealed in the reanalysis data (ERA5). Contrastingly, higher humidity over the equatorial Indian ocean latitudes during the break spell (Fig. 8a) is also captured by all the models (Fig. 8b-e). However, a careful examination can reveal that TaiESM1 (Fig. 8e) showed higher difference in the mid-tropospheric levels that is closer to the reanalysis (Fig. 8a). A similar capability to simulate a higher difference in the mid-tropospheric levels is also noticed in TaiESM1 (Fig. S6e) according to the height-longitude profile of specific humidity difference between active and break spell. IITM-ESM (Fig. S6d) also showed the maxima of differences in the mid tropospheric levels. The vertical profile of difference (i.e., active minus break) averaged over the all-India region (70°E -90°E, 10°N -30°N) also shows that IITM-ESM and TaiESM1 can capture the variation of the difference with the height better than the other two models (Fig. 8f).

**3.1.2.2. High Frequency mode**

The 10-20 days mode or QBM can be considered as a high frequency component of the MISO (Hazra et al., 2019). The dominant westward propagating characteristics of this mode has been studied extensively (Murakami, 1976; Krishnamurti and Ardanuy 1980). The variance of this mode also contributes significantly to the sub-seasonal variance (Goswami, 2012) of the ISMR. The variance of ISMR in QBM scale has been also studied in earlier observational studies (Hazra et al., 2020; Dutta et al., 2020). Over the ISM region, observation (Fig. 9a) shows high variance zone near western Ghats, BoB, and extended to northeastern India. High variances are also noticed over the Central India and, to a less degree, over the EIO region. CanESM5 (Fig. 9b) neglects the high variance zone over major parts of the Indian subcontinent as seen in Observation. FGOALS-g3 (Fig. 9c) also shows variance only over southern peninsula. It also shows high variance zone over the AS and EIO that deviates from the observation. Improvements are noticed for IITM-ESM and TaiESM1. Both can capture the high variance zones near the Western Ghats and over BoB. The variances simulated by these models are also comparable with observation over the EIO (Fig. 9d, e).

To further explore the westward propagation characteristics, we conducted an analysis using lead-lag composites of precipitation filtered over a 10–20-day period relative to a reference time series centered on central India. We present time-longitude sections of these composites averaged within the 10°N–20° N latitude band for the observation and four models (Fig. 9f-j). Hazra et al. (2020) showed that the propagation speed of the QBM specifically over the South China Sea (100° E–120° E), is relatively rapid, with a phase speed of approximately 7.7 m/s. However, as the disturbances move further west into the Bay of Bengal and the Indian monsoon region, the phase speed decreases to around 5.8 m/s (Fig. 9f). Over the South China Sea, CanESM5 (Fig. 9g) contrastingly depicts eastward propagation, whereas FGOALS-g3 (Fig. 9h) shows a sluggish westward propagation. Over the Bay of Bengal regime, the phase speed for CanESM5 is 2.9 m/s, which is much slower than the observed one (Fig. 9b). FGOALS-g3 (Fig. 9h) can simulate the phase speed of 4.63 m/s over BoB. IITM-ESM (Fig. 9i) also simulates the westward propagation at a lower speed (~3.2 m/s). Improvement is shown in TaiESM1 (Fig. 9j) in simulating the phase speed. TaiESM1 simulates the phase speed ~ 5.6 m/s, which is in line with the observation. Hence, the convective coupling over the BoB region is considerably well captured in TaiESM1 than the other three models.

### 3.1.2.3. Low-Frequency Mode

To investigate the performance of the four models on depicting the low frequency mode of the MISO, we present north–south space–time spectra of zonally averaged daily rainfall anomalies (June–September) within the region from 60° E to 110° E, filtered over a period of 20 to 100 days. The methodology used for computation is based on the approach outlined by Wheeler and Kiladis (1999). In this context, the meridional wavenumber 1 corresponds to the largest wave that can fit within the latitudinal extent ranging from 20° S to 33° N, as described by Goswami et al. (2011). On the horizontal axis, the positive values indicating northward propagation and negative values representing southward propagation. Previous studies (e.g., Dutta et al., 2020, 2021) have shown a predominant mode of northward-propagating ISO, located at wavenumber 1 with a periodicity of 40 days. This mode is more pronounced than its southward counterpart (Fig. 10a). The CanESM5 (Fig. 10b) unrealistically shows the northward and southward propagation with similar intensity, while the other three models have shown improvement in this regard. FGOALS-g3 (Fig. 10c) capture the maximum intensity at wavenumber 2 and periodicity around 30 days. Though IITM-ESM (Fig. 10d) simulates the intensity less than that of observation, it can capture the maxima at wavenumber 1 and periodicity of 40 days similar to the observation. TaiESM1 simulated the maxima of northward propagation but with a wavenumber between 1 and 2 and periodicity around 30 days. Nevertheless, the magnitude of the maximum power of the northward propagation is better simulated in TaiESM1 (Fig. 10e) among all the models.

One of the most significant features of MISO is the northward propagation of the 30–60-day oscillations of convection anomalies (Sharmila et al., 2013; Hazra et al., 2017a). The intraseasonal variance of rainfall of this band is shown from observation and models (Fig. 11a-e). The observation (Fig. 11a) depicts a zone of high variance over the West coast of India and adjacent the Arabian Sea, northern Bay of Bengal, and the South China Sea. CanESM5 fails to replicate the substantial variability observed over a significant portion of the Indian subcontinent (Fig. 11b). Similarly, FGOALS-g3 exhibits variance primarily over the southern peninsula, the Arabian Sea, and the Eastern Indian Ocean (EIO), which deviate from the observed pattern. In contrast, IITM-ESM and TaiESM1 demonstrated enhancements in this regard. Both models successfully reproduce the high-variance zones near the western Ghats and over the Bay of Bengal (BoB), aligning more closely with the observed patterns. Over the central India, simulated variance is better represented in TaiESM1 (Fig. 11e) than IITM-ESM (Fig. 11d).

Lag-Latitude (averaged over 70° E–90° E) diagram (e.g., Saha et al. 2013) of regressed (lead/lag) 30–60 days filtered rainfall anomaly with reference to central India time series is shown from observation and four models (Fig. 11f-j). The observation shows northward propagation speed around 1.58°/day. CanESM5 shows a stationary wave over the ISM region. Among the four models IITM-ESM shows the propagation speed of approximately 1.25°/day. However, all the four models (Fig. 11f-j) cannot realistically simulate the northward propagation.

### 3.2 Madden–Julian Oscillation (MJO) Features:

### 3.2.1. The Mean State of MJO

We have performed the analyses as per the standard MJO diagnostics, developed by the CLIVAR MJO Working Group (Waliser et al., 2009). To analyze MJO features, first, the bias of mean rainfall from four models for both boreal summer (May to October; MJJASO) and boreal winter (November to April; NDJFMA) are presented in Figure 12. During boreal summer (Fig. 12a-d), CanESM5 and FGOALS-g3 both show much dry bias over the Indian landmass. Over the China CanESM5 (FGOALS-g3) show wet (dry) bias. Over the south China sea (SCS) CanESM5 show wet bias which is significantly reduced in TaiESM1 (Fig. 12d). IITM-ESM (Fig. 12c) and FGOALS-g3 (Fig. 12b) also show less wet bias over the SCS. Over the equatorial Pacific Ocean CanESM5 shows dry bias with wet bias north and south of it. Other models mostly show wet bias over the equatorial Pacific Ocean. Over the north (south) equatorial Atlantic Ocean FGOALS-g3 and IITM-ESM show dry (wet) bias. The bias is also improved in TaiESM1. Over the Central Africa region also the bias is improved in TaiESM1 (Fig. 12d). Quantitative bias estimate over the global tropics (0-360, 30°S-30°N) shows that the all the four models show wet bias for boreal summer. The magnitude of bias is the highest in TaiESM1 (~ 0.64 mm/day) and the lowest in FGOALS-g3 (~ 0.2 mm/day). IITM-ESM (~ 0.24 mm/day) shows less wet bias than CanESM5 (~ 0.5 mm/day).

During boreal winter (Fig. 12e-h), the wet bias is shown over the majority area of the southern Indian ocean by all the models. However, the magnitude is reduced in IITM-ESM (Fig, 12g) and TaiESM1(Fig. 12h) as compared to CanESM5 (Fig. 12e) and FGOALS-g3 (Fig. 12f). High wet bias in the southern Atlantic Ocean is noticed in IITM-ESM. Over the Latin America continental landmass, the dry bias is much reduced in TaiESM1. Quantitative estimate for boreal winter over global tropics shows that all the models show wet bias. IITM-ESM (Fig. 12g) shows the least wet bias (~ 0.25 mm/day) whereas TaiESM1 (Fig. 12h) shows the highest wet bias (~ 0.7 mm/day). CanESM5 and FGOALS-g3 show wet bias of approximately 0.5 mm/day and 0.3 mm/day respectively (Fig. 12 a, b). The OLR biases (Figure not shown) are also in agreement with the distribution of rainfall bias. During boreal summer overestimation of OLR (i.e., suppression of convection) are shown over the major parts of the Indian subcontinent by all the models. However, in terms of magnitude, distinct improvement is seen in IITM-ESM and TaiESM1. The overestimation of OLR over the Latin America continental landmass during boreal summer is also significantly reduced in TaiESM1 which is also in agreement in reduction of dry bias over this region among the models.

### 3.2.2. Intraseasonal Features:

The MJO is a characterized by an eastward moving (average speed of ~ 5 m/s) large scale convection (Li et al., 2022) from the Indian Ocean to the central Pacific (Jiang et al., 2015). However, due to barrier effect (Hsu and Lee, 2005; Rui and Wang, 1990) the MJO gets weakened over the maritime continents and fails to propagate further toward western Pacific. The MJO also have seasonal variation of intensity (Zhang and Dong, 2004). Over the western Pacific the MJO is stronger during winter and weaker in summer (Chen et al., 2017). Hung et al., (2013) found that most of the climate models fail to capture these features of MJO. We have analyzed the four models regarding these intraseasonal features of the MJO. We have first computed the Lag-longitude diagram of the observations and four models for both the seasons i.e., boreal summer (Fig. 13) and winter (Fig. 14), which is considered as an important diagnostic for testing the fidelity of the models (Waliser et al., 2009, Dutta et al., 2021, Ahn et al., 2020). Following the conventional methodology (e.g., Li et al., 2022; Ahn et al., 2020) the intraseasonal (20–100 days) precipitation anomalies are first regressed against a reference Indian Ocean (85°E–95°E and 5°S to 5°N) intraseasonal precipitation time series. In the next step, we consider the meridional average over the latitudes of 10°S to 10°N.

During the boreal summer, observation shows continuous eastward propagation from 60°E towards the vicinity of 180°E (Fig. 13a). CanESM5 poorly captures this propagation feature and almost shows stationary waves around the convection maxima (Fig. 13b). FGOALS-g3 (Fig. 13c) performs better than CanESM5 and IITM-ESM (Fig. 13d) in this regard. Interesting improvement is noticed in TaiESM1 (Fig. 13e) which can capture the observation more realistically among the four models. Recently Ahn et al., (2020) has defined a metric to evaluate the fidelity of the models in depicting the propagation of MJO across the maritime continents. Following the methodology, we have averaged the regression coefficients over the domain 100°E-150°E and duration 0 to 25 days from the lag-longitude diagram for the four models, further divided by the same that of the observation. CanESM5 (~0.46) and IITM-ESM (~0.45) both yield a very low value (i.e., less than 0.5). FGOALS-g3 is closer to observation with metric value of 0.95. TaiESM1 on the other hand shows higher metric value of 1.89. During boreal winter (Fig. 14) the speed of propagation is faster than that of summer and the barrier effect near the maritime continents (near 120°E) is more pronounced in observation (Fig. 14a).

Likewise, summer, the CanESM5 also sparsely simulates the eastward propagation (Fig. 14b) which is improved in the other three models. IITM ESM shows the eastward propagation from around 60°E, but fails to show continuous propagation till 180°E (Fig. 13d) as seen from the observation (Fig. 14a), which is improved in FGOALS-g3 (Fig. 14c) and TaiESM1 (Fig. 14e). The barrier effect near the maritime continents is also seen in TaiESM1. CanESM5 shows very less value (~ 0.2) of propagation metric, whereas the other three models (FGOALS-g3: 0.63; TaiESM1: 0.72; IITM-ESM: 0.92) show well value (i.e., greater than 0.5), in which IITM-ESM is the closest to the observation value.

The wavenumber-frequency power spectra are applied on rainfall anomalies averaged over 10°S-10°N to isolate the spatial and temporal signals which are related to MJO as discussed in earlier studies (Jiang et al., 2022, Li et al., 2022) for boreal summer (Fig. S7), and winter (Fig. S8) seasons from the observation and CMIP6 models. Observation shows dominant eastward propagation for both the seasons. For summer, the eastward propagation is observed (Fig. S7a) between zonal wavenumber 1-3 and the maxima is located at wavenumber 1. The maximum of eastward propagation is also found between 30-80 days of periodicity. In case of models, CanESM5 (Fig.S7b) fails to capture the propagation feature. FGOALS-g3 can capture the eastward propagation spectra between zonal wavenumber 1-4 and with periodicity between 30-80 days (Fig. S7c), which is comparable with observation. However, it cannot show the peak at zonal wavenumber 1 like observation. TaiESM1 (Fig. S7e) can capture the peak at zonal wavenumber 1 and the propagation of 30-80 days periodicity, but the magnitude of the power is diminished. IITM-ESM (Fig. S7d) fails to simulate the dominant eastward propagation. For boreal winter also the eastward propagation is dominating than its westward counterpart as seen from the observation (Fig. S8a). The propagation is seen between zonal wavenumber 1-3 with periodicity between 25-100 days. All the models cannot properly simulate the propagation feature. IITM-ESM here can at least capture the dominant eastward propagation (Fig. S8d).

## 4. Summary

This study assessed the performance of four climate models in capturing various features related to monsoon rainfall, including mean state, intraseasonal oscillations (MISO), and Madden-Julian Oscillation (MJO). The key findings from this study are summarized below:

1. All four models overestimated the mean value of global tropics rainfall during the Indian summer monsoon (JJAS). The models exhibit differences in area-wise distribution over various parts of the tropics, with TaiESM1 performing best over the East Indian Monsoon Region (EIMR), West Pacific, and Amazon region. The models generally capture the observed convection patterns, with TaiESM1 being closest to observations, but all models tend to overestimate OLR over the global tropics. TaiESM1 shows the highest mean bias in OLR over the global tropics, while IITM-ESM has the lowest bias. The spatial distribution of TCF also varies among the models, with TaiESM1 showing improvements over CanESM5 and FGOALS-g3.

2. All models have a common issue of dry bias over the Indian subcontinent during monsoon, but the intensity and spatial extent vary. TaiESM1 performed better, showing the least dry bias among the four models over central India, and exhibiting improved bias over the equatorial Indian Ocean and the Bay of Bengal. TaiESM1 also exhibits less bias in TCF over the South Asian monsoon region as compared to other models.

3. Active-break spells during the monsoon are crucial for the ISM, and TaiESM1 shows better representation than CanESM5 and FGOALS-g3. TaiESM1 and IITM-ESM demonstrated improvements in capturing the northward-propagating, 30–60-day MISO over the Indian subcontinent compared to CanESM5 and FGOALS-g3. The quasi-bi-weekly mode (QBM) is better simulated in TaiESM1, with improvements in capturing variance over the Indian subcontinent.

4. Mean state analysis during boreal summer and winter for MJO indicates that TaiESM1 performs comparatively better than CanESM5, FGOALS-g3, and IITM-ESM. TaiESM1 shows improvements in capturing eastward propagation during both summer and winter, with a better representation of the barrier effect over maritime continents.

Hence, the results from the study show that TaiESM1 and IIT-ESM which has improved the mean representation of ISM also provides better fidelity in simulating the MISO and MJO features. The improvements in the simulations can be attributed to the better representation of clouds and convection. We hope that the understanding from this study will help the researchers for improving the ISM prediction.

**Acknowledgements:** We acknowledge the climate modeling groups for providing their model output via World Climate Research Program, the Earth System Grid Federation (ESGF). We also thank the freely available software viz. The Grid Analysis and Display System (GrADS), NCAR Command Language (NCL), Ferret-NOAA, Climate Data Operators (CDO), and Python (https://www.python.org/) for producing the results. The first author (UD) thanks the funding support received from National Science and Technology Council (NSTC) – Taiwan. The second and third author (MB and AH) thank MoES, Government of India, and Director IITM. AH also dully acknowledges National Science and Technology Council (NSTC), Taiwan for funding support as visiting researcher in National Taiwan University, Taiwan. The authors declare that they have no known competing financial interests or personal relationships that could have appeared to influence the work reported in this paper.

**Data Availability Statement:** All data used in this study are freely available in the public domain. The Coupled Model Inter-comparison Projects (CMIP6) data are collected from World Climate Research Programm and available at the Earth System Grid Federation (ESGF, https://esgf-node.llnl.gov/). The fifth generation of the European Centre for Medium-Range Weather Forecasts (ECMWF) reanalysis, ERA5 (Hersbach et al., 2020) data are available at (https://cds.climate.copernicus.eu/#!/search?text=ERA5&type=dataset). The OLR data are used from https://psl.noaa.gov/data/gridded/data.interp_OLR.html (Liebmann and Smith, 1996).The GPCP rainfall data are available at (http://www.esrl.noaa.gov/psd/data/gridded/data.gpcp.html, Adler et al., 2003).

**List of Figures**

**Figure 1:** Bias of total cloud fraction (TCF, %) and rainfall (mm/day) over the central India region from several CMIP6 models.

**Figure 2:** June to September (JJAS) climatology of a) rainfall (mm/day) b) outgoing longwave radiation (OLR; Watt/m$^2$) and c) total cloud fraction (TCF; %) over the global tropics from observation/reanalysis.

**Figure 3:** Rainfall bias of the selected four models (a-d) with respect to the observation (GPCP) during JJAS season over the south Asian monsoon region. e) Probability density of the daily rainfall over central India region from observation and four models. All India box region (70˚E -90˚E, 10˚N -30˚N) is shown in box in (b).

**Figure 4:** OLR bias of the selected four models (a-d) with respect to the observation (GPCP) during JJAS season over the south Asian monsoon region. e) Probability density of the daily OLR over EIMR from observation (NOAA) and four models.

**Figure 5:** TCF bias (%) of the selected four models (a-d) with respect to the reanalysis (ERA5) during JJAS season over the south Asian monsoon region. Average bias of TCF over the EIMR (shown is box in all panels) is written in parenthesis over each panel.

**Figure 6:** Difference between active and break spell of rainfall (mm/day) from observation (a) and four models (b-e).

**Figure 7:** Same as Figure 6 but for total cloud fraction (TCF; %).

**Figure 8**: Height-Latitude distribution of difference between active and break spell of specific humidity (g/kg) averaged over longitude (70˚E-90˚E) from reanalysis-ERA5 (a) and four models (b-e). Vertical profile of similar difference of specific humidity over all India region (70˚E -90˚E, 10˚N -30˚N) in shown in (f).

**Figure 9:** Averaged variance (mm$^2$/day$^2$) of high-frequency (10-20 days band) component of monsoon rainfall from observation (a) and four models (b-e). Lag-longitude regression of rainfall (averaged over 10°N-20°N) with respect to time series of central India region from observation (f) and four models (g-j).

**Figure 10:** North-South Spectra of 60°E–110°E averaged intraseasonal (20–100 days) rainfall anomalies from observation-GPCP (a) and four models (b-e). Positive (Negative) frequency denotes northward (southward) propagation.

**Figure 11:** Averaged variance ($mm^2/day^2$) of low-frequency (30-60 days band) component of monsoon rainfall from observation (a) and four models (b-e). Lag-latitude regression of filtered (30-60 days band) rainfall (averaged over 70°E-90°E) anomalies with respect to time series of central India region from observation (f) and four models (g-j).

**Figure 12:** Rainfall bias (mm/day) of four models (a-d) with respect to observation (GPCP) for boreal summer season (May-Oct). Similar bias but for boreal winter season (Nov-Apr) of four models are also shown (e-h).

**Figure 13:** Lag-longitude regression of filtered (20-100 days band) rainfall (averaged over 10°S-10°N) anomalies with respect to time series of Indian ocean (85°E–95°E and 5°S-5°N) region from observation (a) and four models (b-e) during boreal summer season (May-Oct).

**Figure 14:** Same as Figure 13 but for boreal winter (Nov-Apr) season.

**List of Tables:**

**Table 1.** Biases of rainfall (mm/day) over different parts of the south Asian monsoon region (SAMR). Positive(negative) value signifies wet (dry) bias.

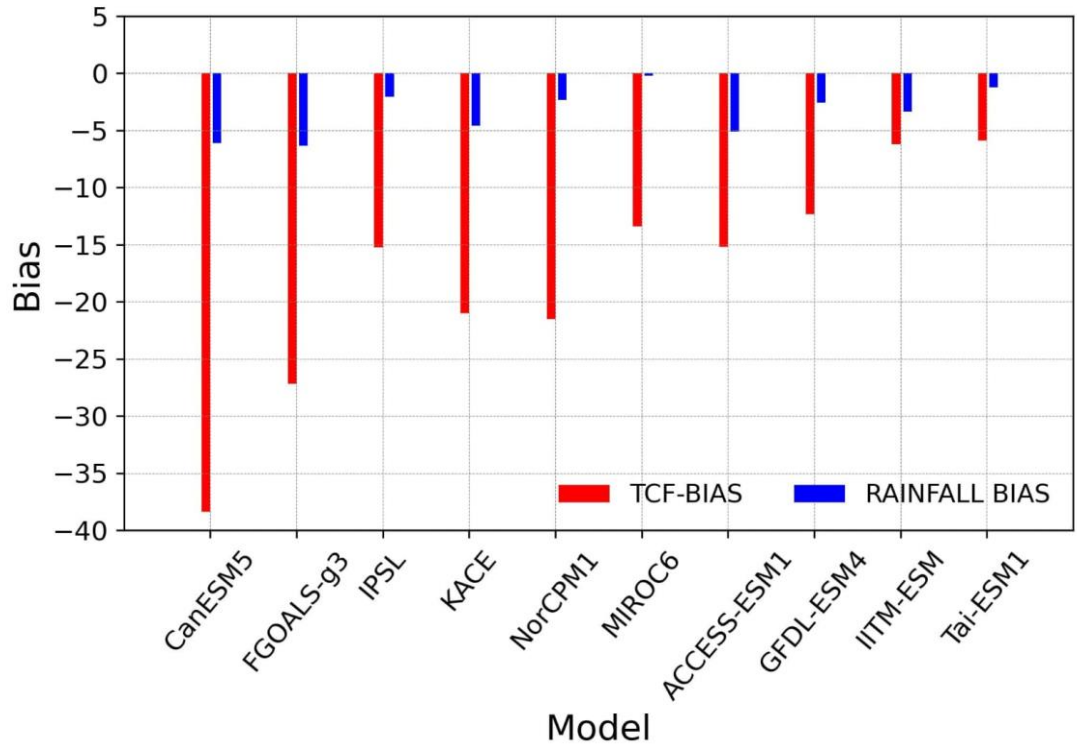

**Figure 1:** Bias of total cloud fraction (TCF, %) and rainfall (mm/day) over the central India region from several CMIP6 models.

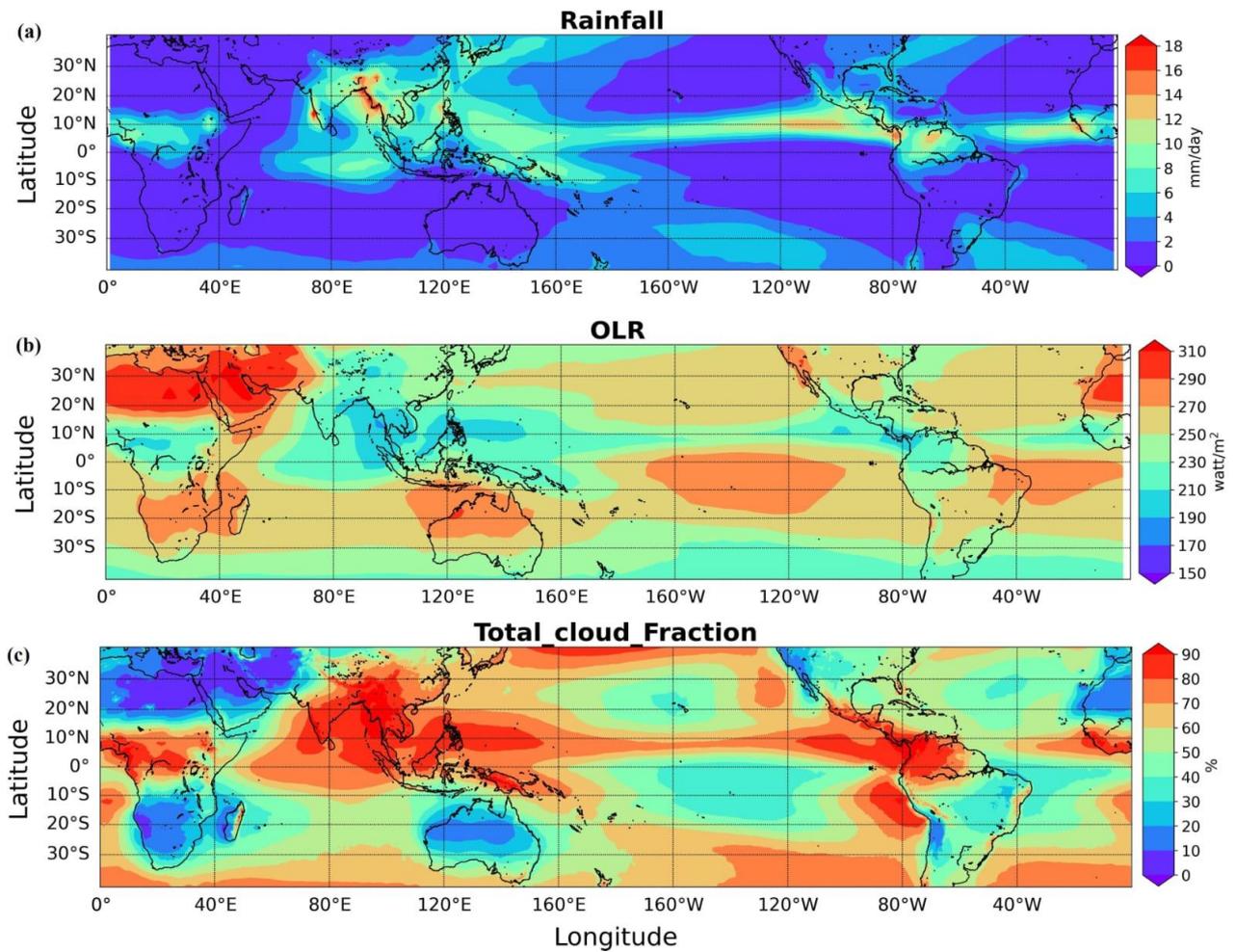

**Figure 2:** June to September (JJAS) climatology of a) rainfall (mm/day) b) outgoing longwave radiation (OLR; Watt/m$^2$) and c) total cloud fraction (TCF; %) over the global tropics from observation/reanalysis.

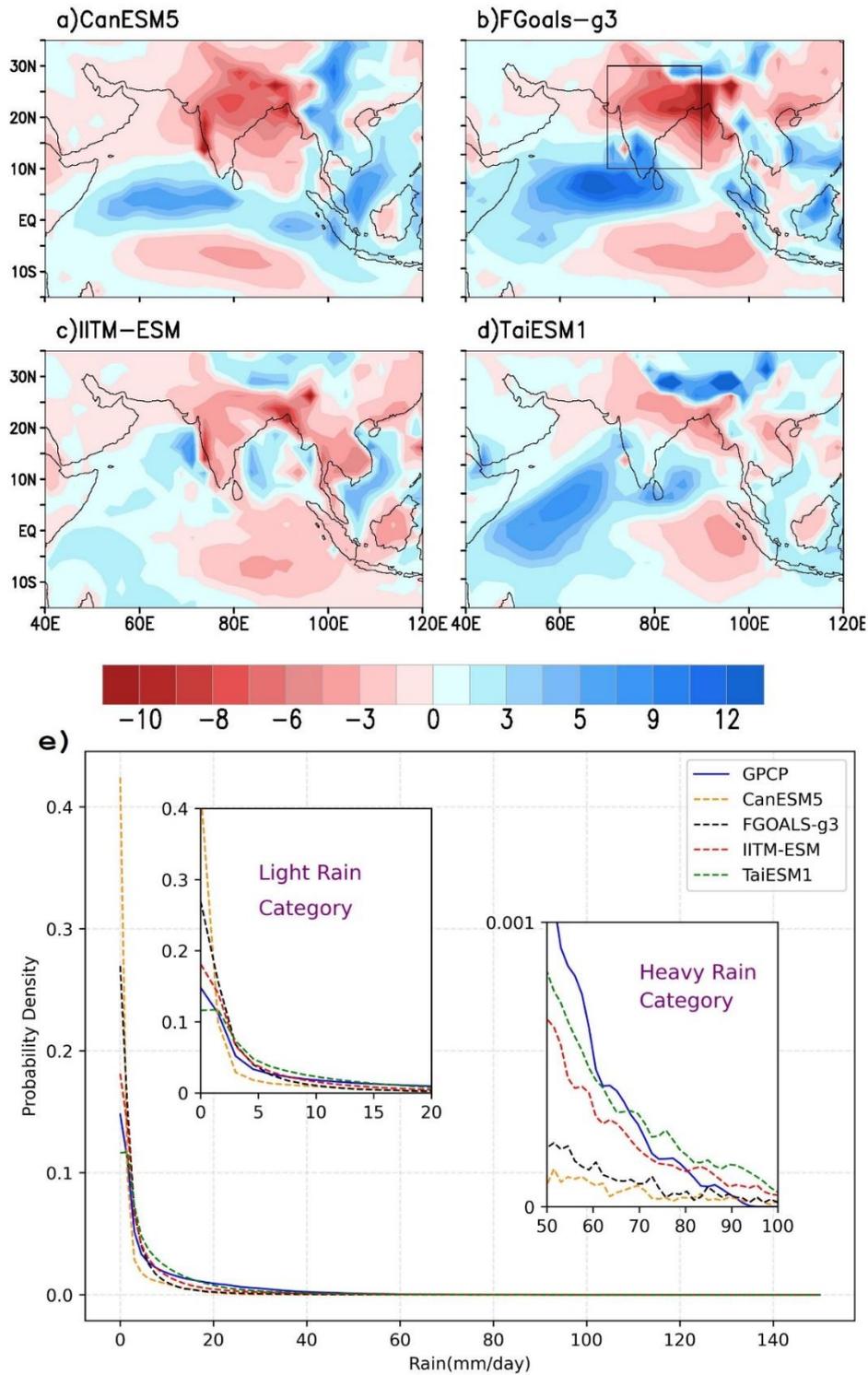

**Figure 3:** Rainfall bias of the selected four models (a-d) with respect to the observation (GPCP) during JJAS season over the south Asian monsoon region. e) Probability density of the daily rainfall over central India region from observation and four models. All India box region (70°E -90°E, 10°N -30°N) is shown in box in (b).

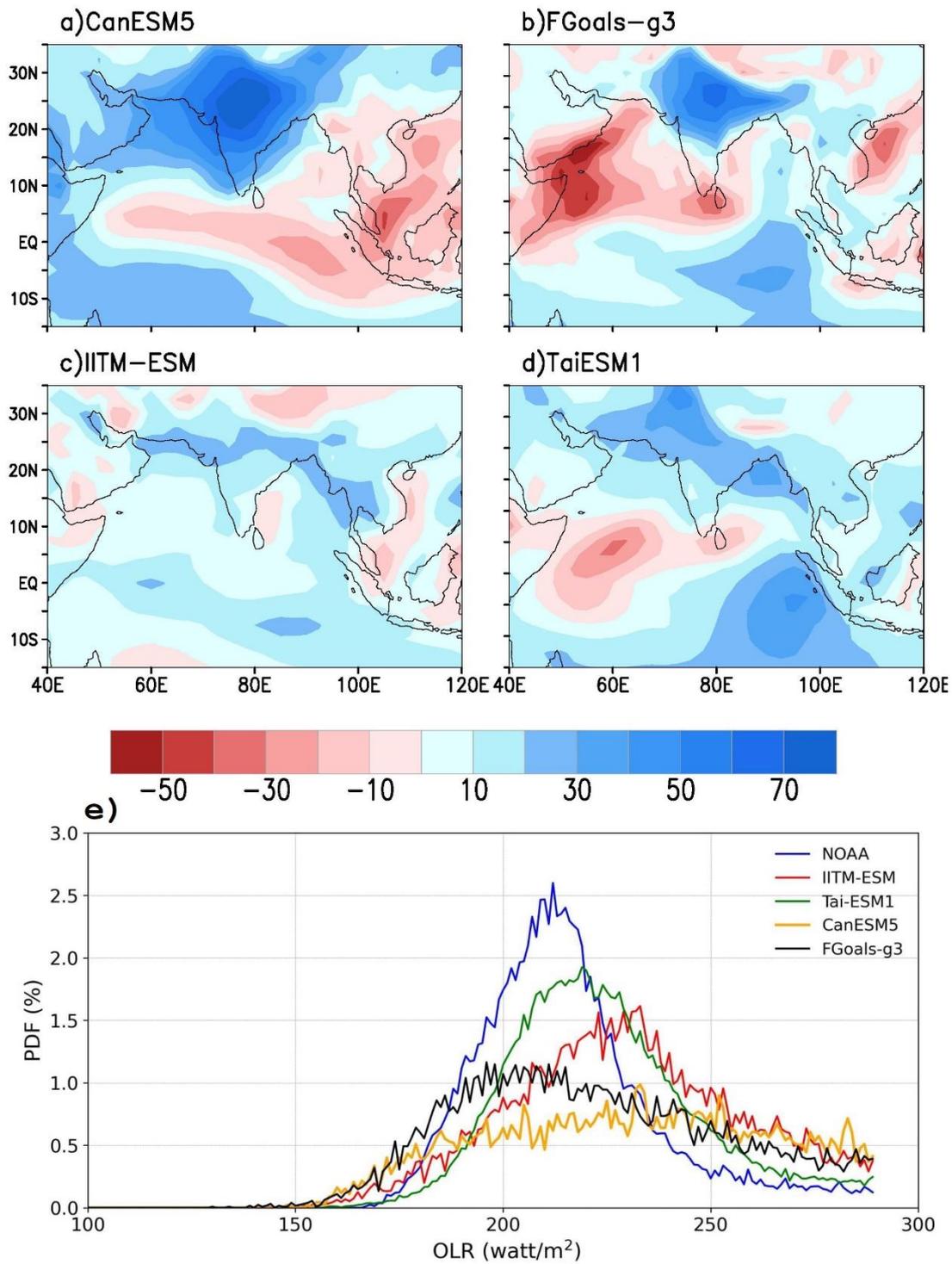

**Figure 4:** OLR bias of the selected four models (a-d) with respect to the observation (GPCP) during JJAS season over the south Asian monsoon region. e) Probability density of the daily OLR over EIMR from observation (NOAA) and four models.

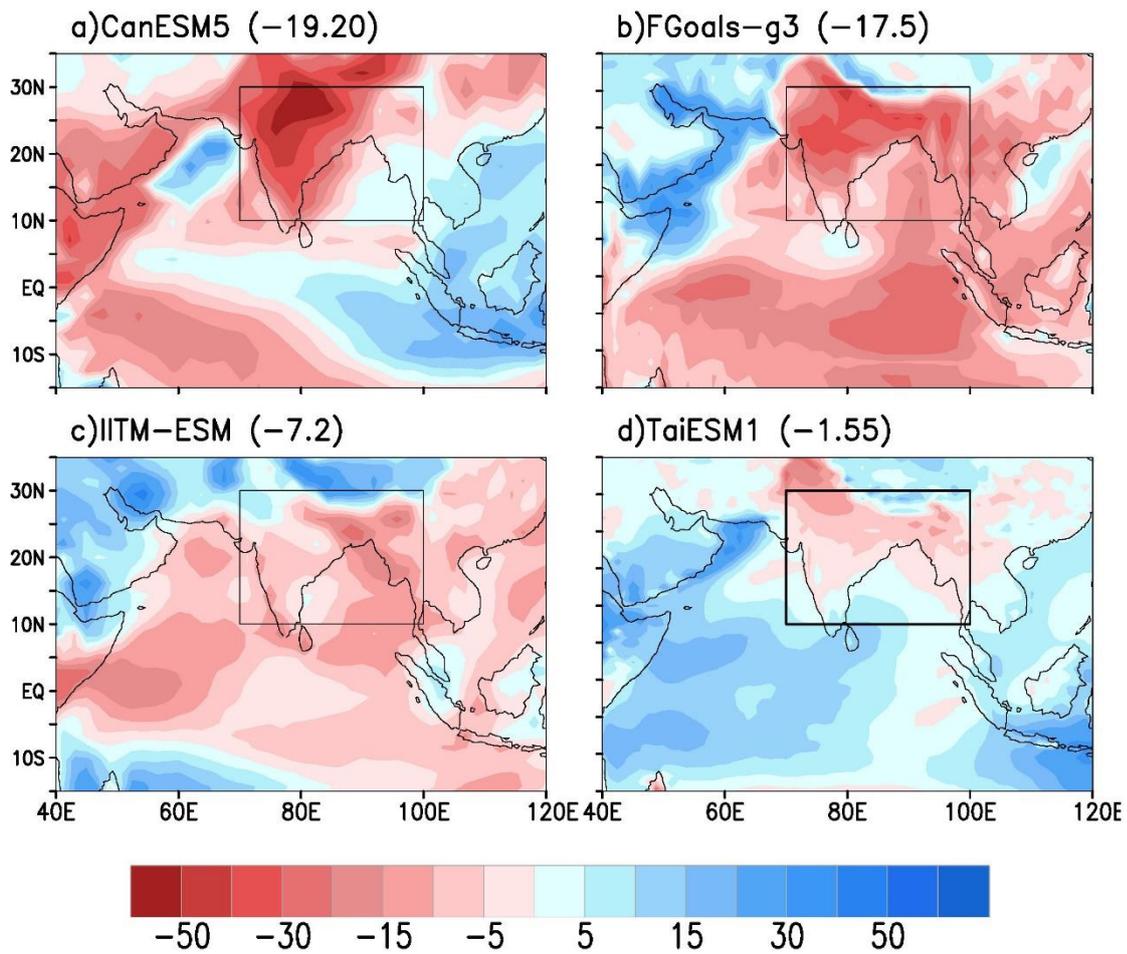

**Figure 5:** TCF bias (%) of the selected four models (a-d) with respect to the reanalysis (ERA5) during JJAS season over the south Asian monsoon region. Average bias of TCF over the EIMR (shown is box in all panels) is written in parenthesis over each panel.

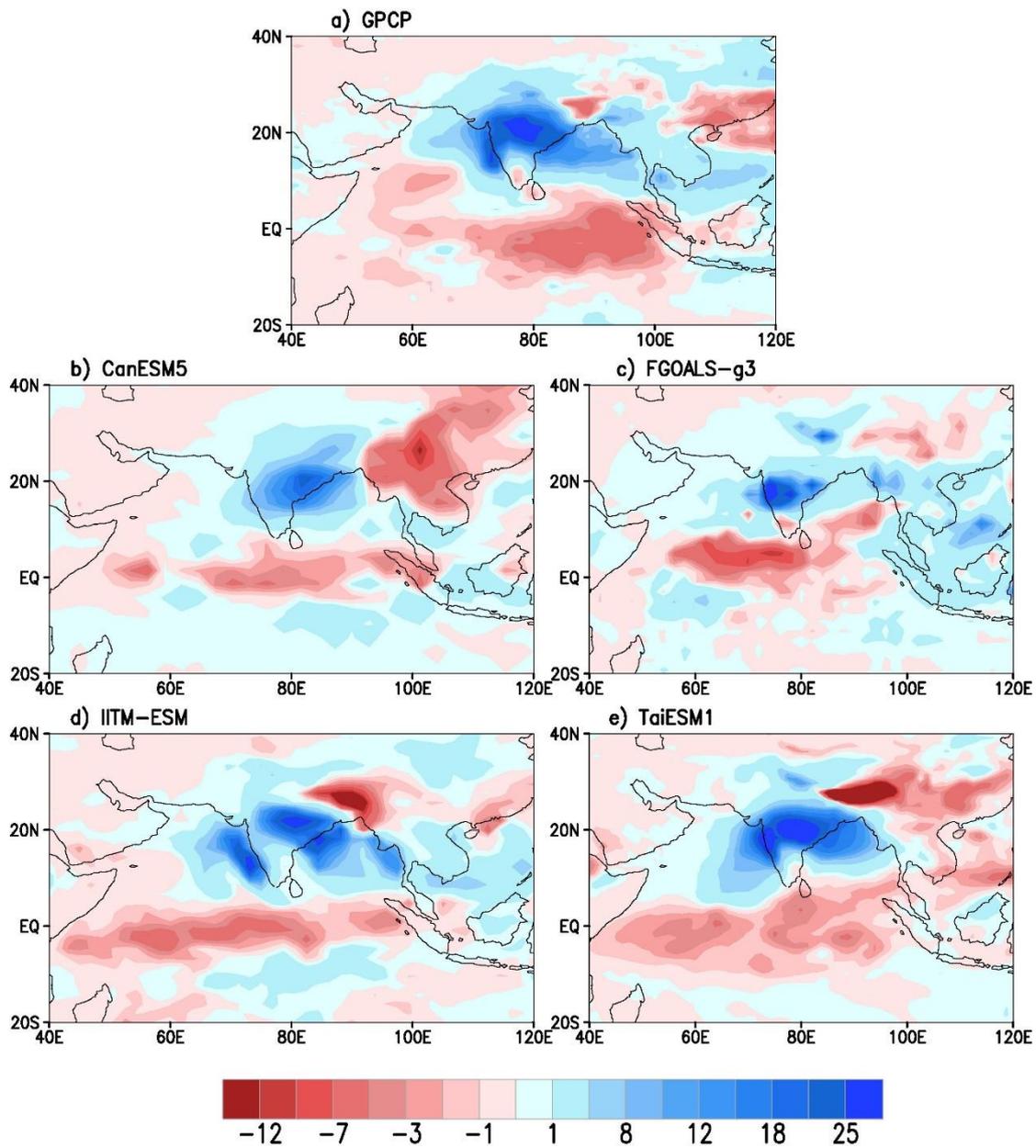

**Figure 6:** Difference between active and break spell of rainfall (mm/day) from observation (a) and four models (b-e).

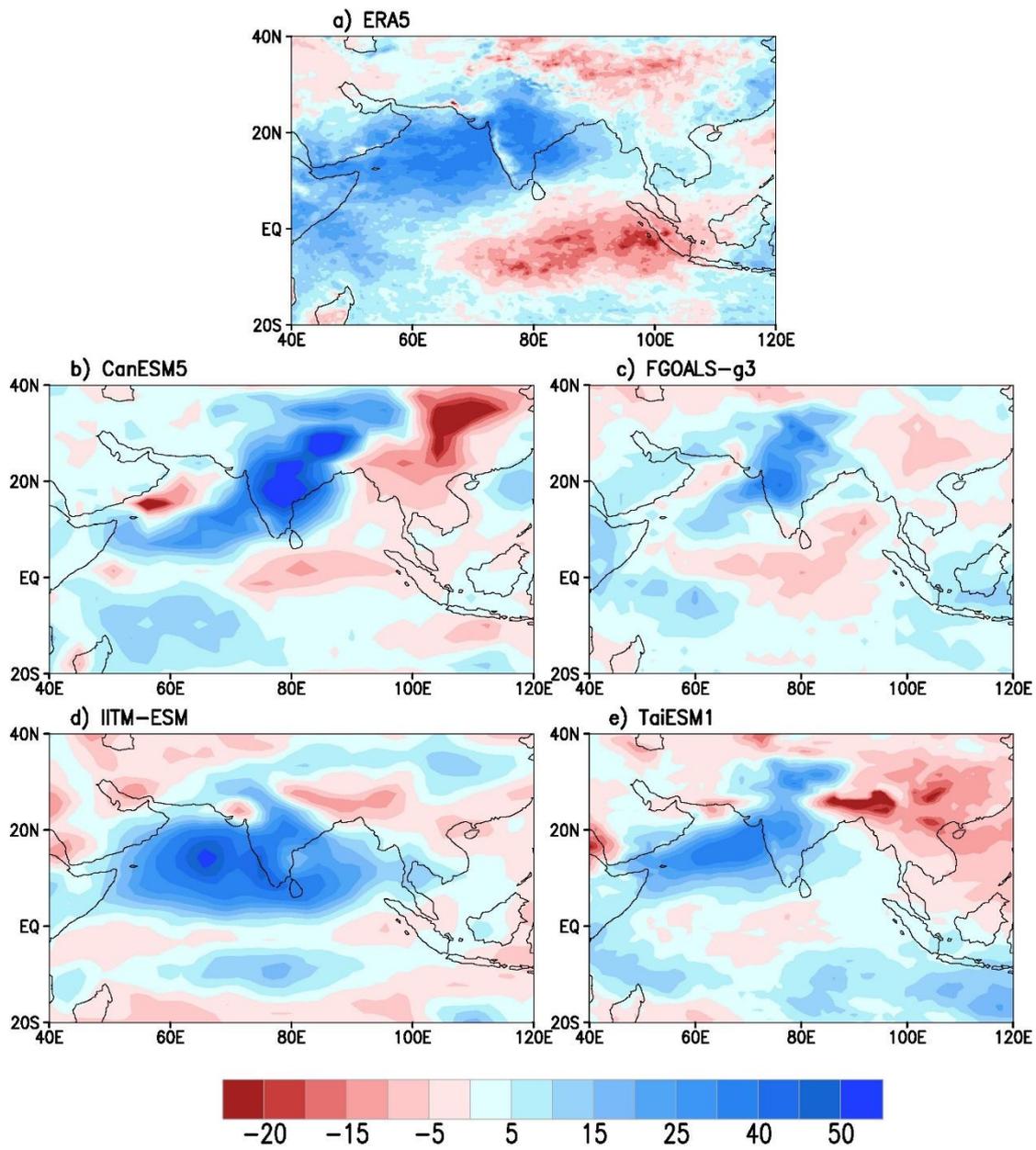

**Figure 7:** Same as Figure 6 but for total cloud fraction (TCF; %)

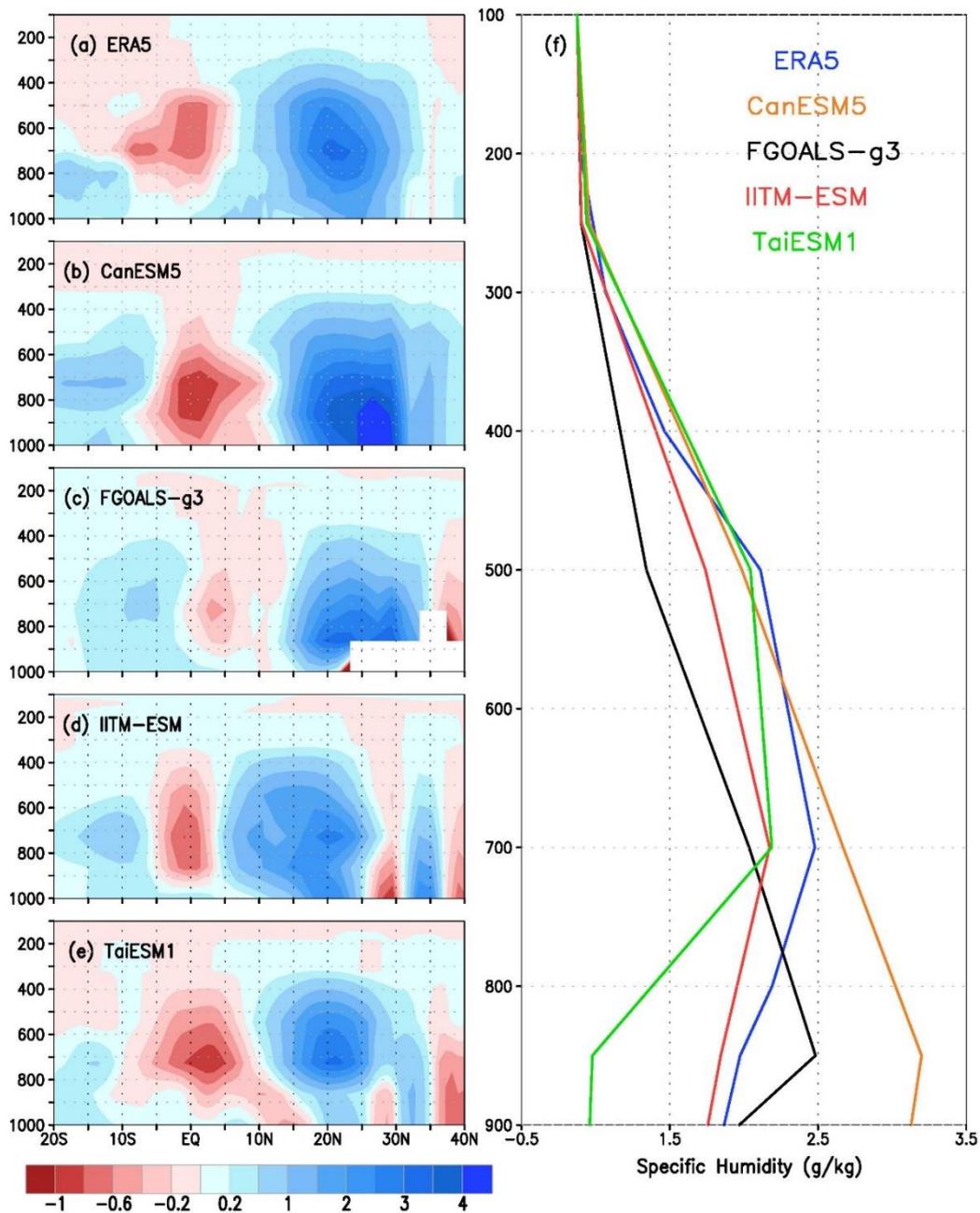

**Figure 8:** Height-Latitude distribution of difference between active and break spell of specific humidity (g/kg) averaged over longitude (70°E-90°E) from reanalysis-ERA5 (a) and four models (b-e). Vertical profile of similar difference of specific humidity over all India region (70°E -90°E, 10°N -30°N) in shown in (f).

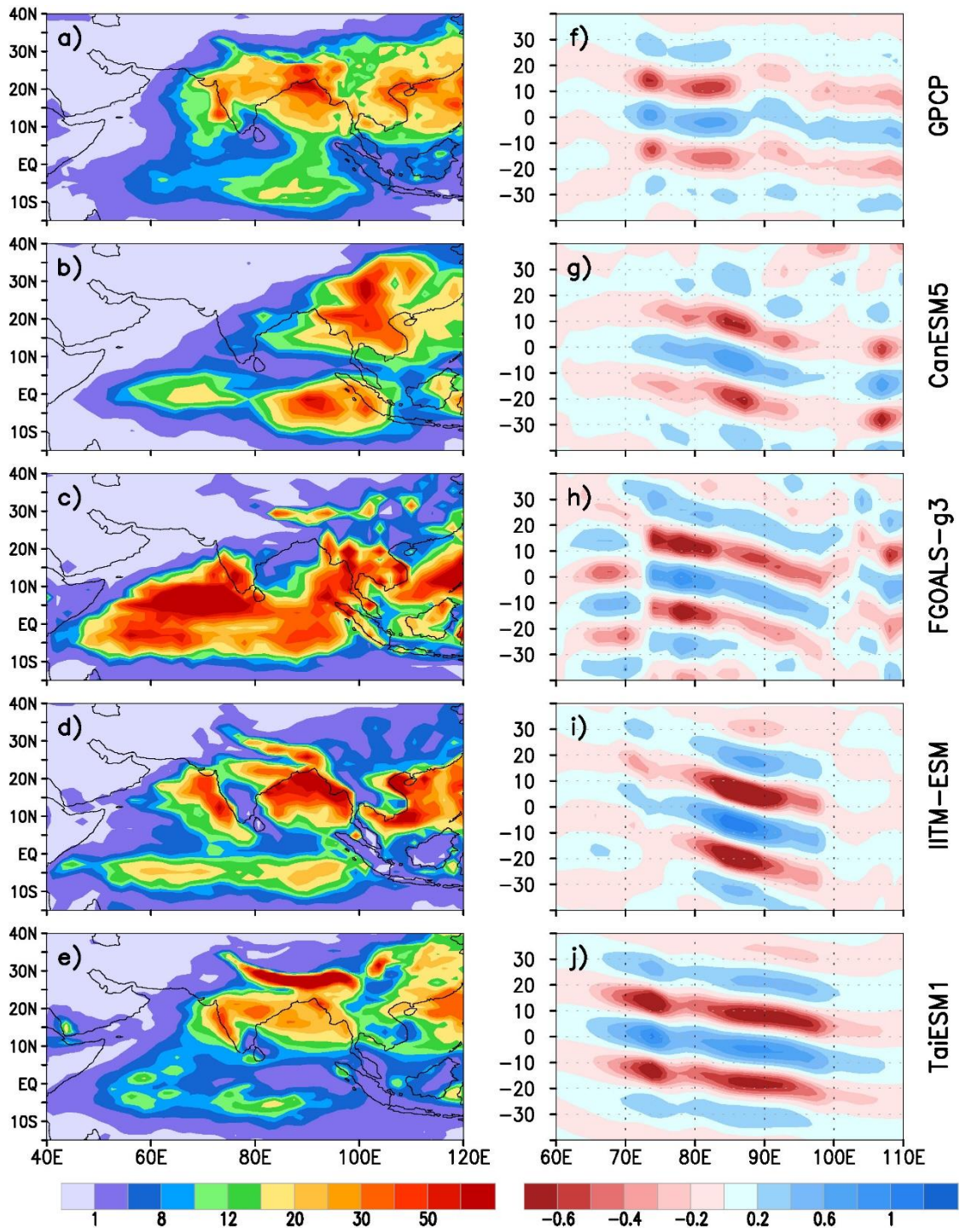

**Figure 9:** Averaged variance (mm$^2$/day$^2$) of high-frequency (10-20 days band) component of monsoon rainfall from observation (a) and four models (b-e). Lag-longitude regression of rainfall (averaged over 10°N-20°N) with respect to time series of central India region from observation (f) and four models (g-j).

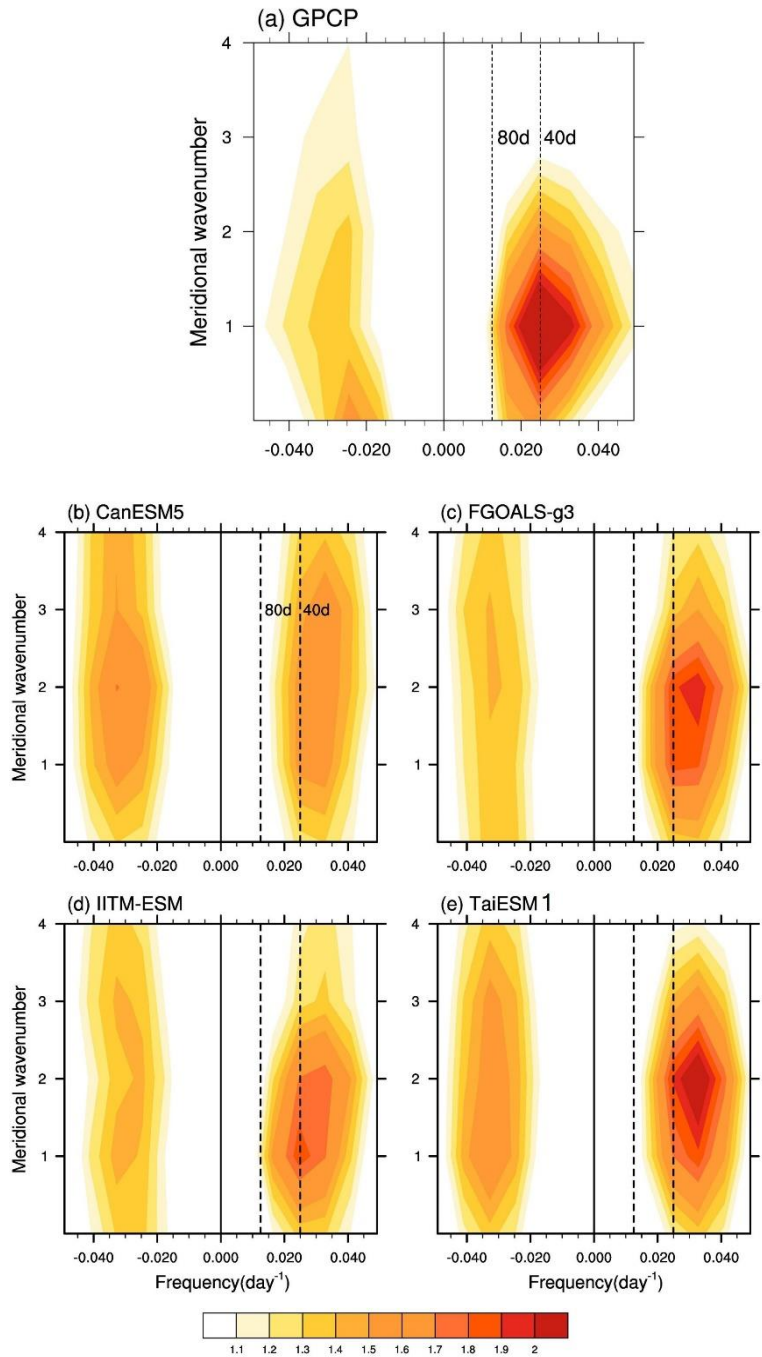

**Figure 10:** North-South Spectra of 60°E–110°E averaged intraseasonal (20–100 days) rainfall anomalies from observation-GPCP (a) and four models (b-e). Positive (Negative) frequency denotes northward (southward) propagation.

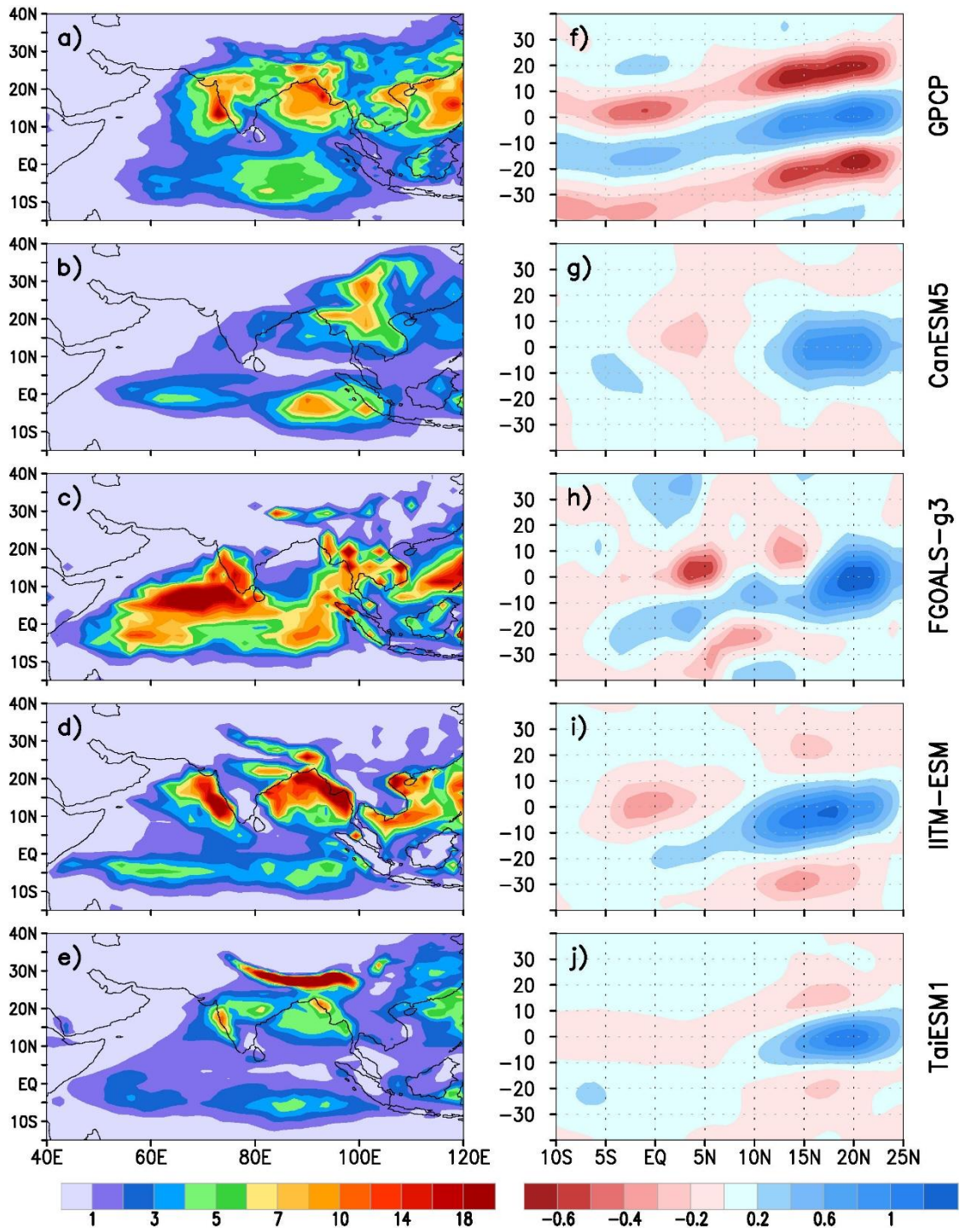

**Figure 11:** Averaged variance (mm$^2$/day$^2$) of low-frequency (30-60 days band) component of monsoon rainfall from observation (a) and four models (b-e). Lag-latitude regression of filtered (30-60 days band) rainfall (averaged over 70°E-90°E) anomalies with respect to time series of central India region from observation (f) and four models (g-j).

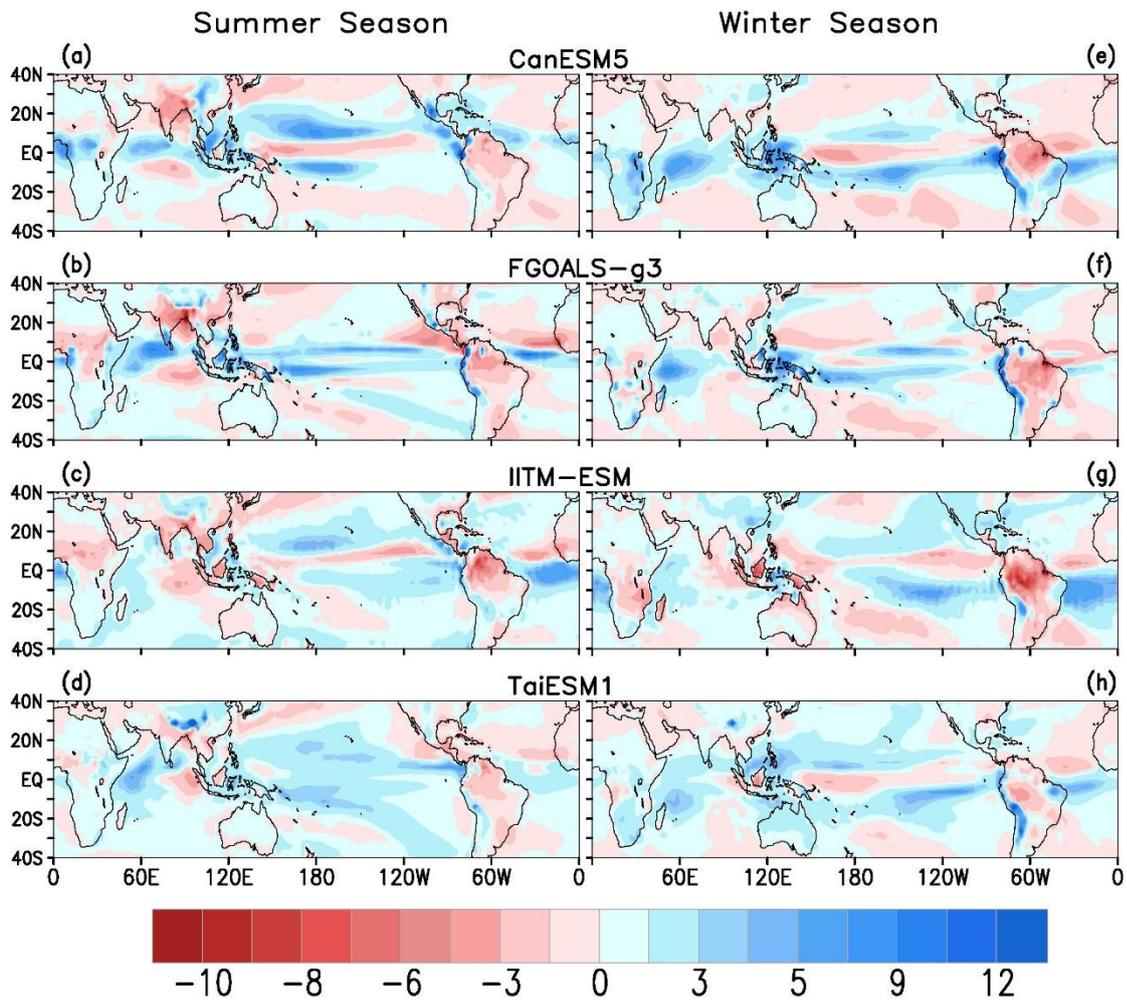

**Figure 12:** Rainfall bias (mm/day) of four models (a-d) with respect to observation (GPCP) for boreal summer season (May-Oct). Similar bias but for boreal winter season (Nov-Apr) of four models are also shown (e-h).

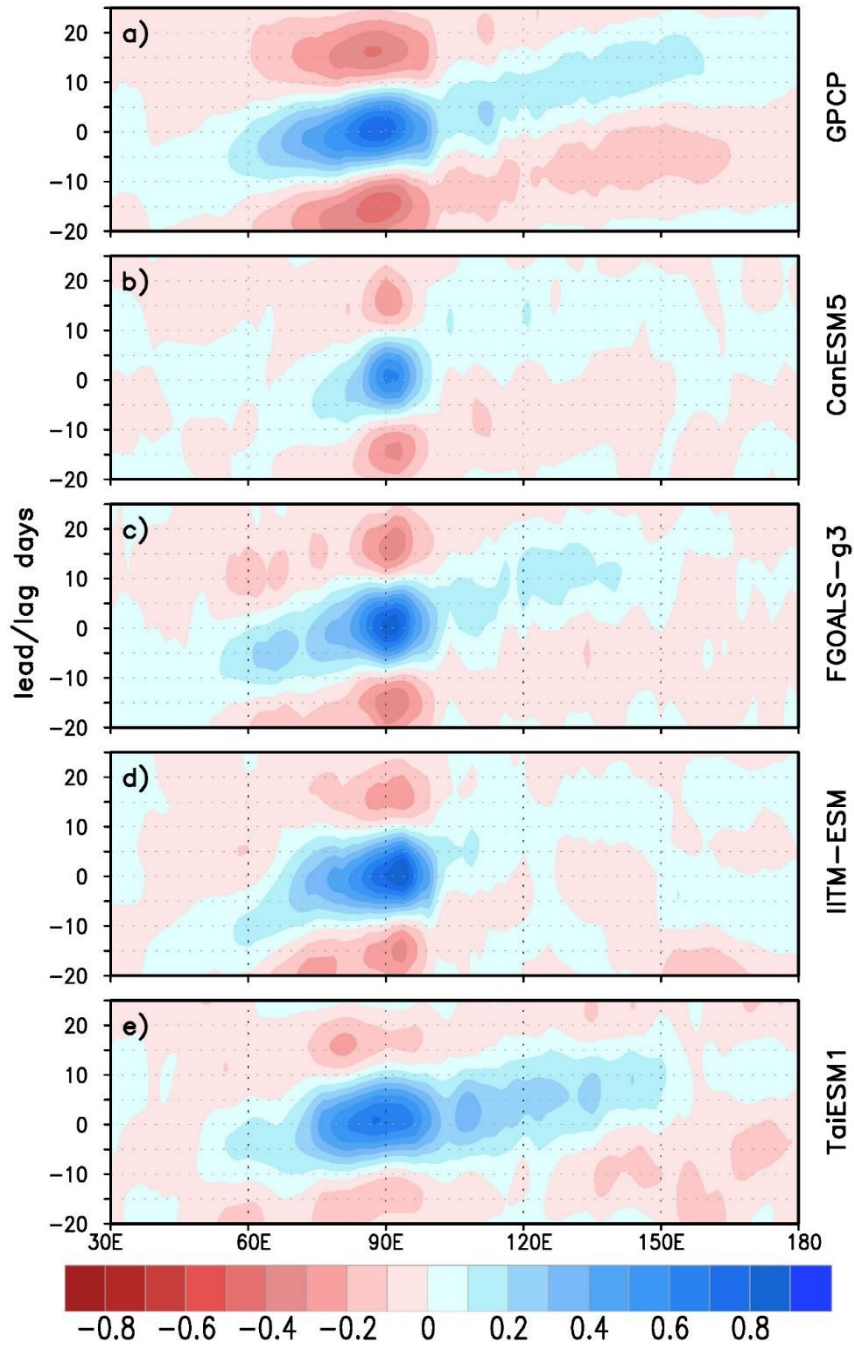

**Figure 13:** Lag-longitude regression of filtered (20-100 days band) rainfall (averaged over 10°S-10°N) anomalies with respect to time series of Indian ocean (85°E–95°E and 5°S-5°N) region from observation (a) and four models (b-e) during boreal summer season (May-Oct).

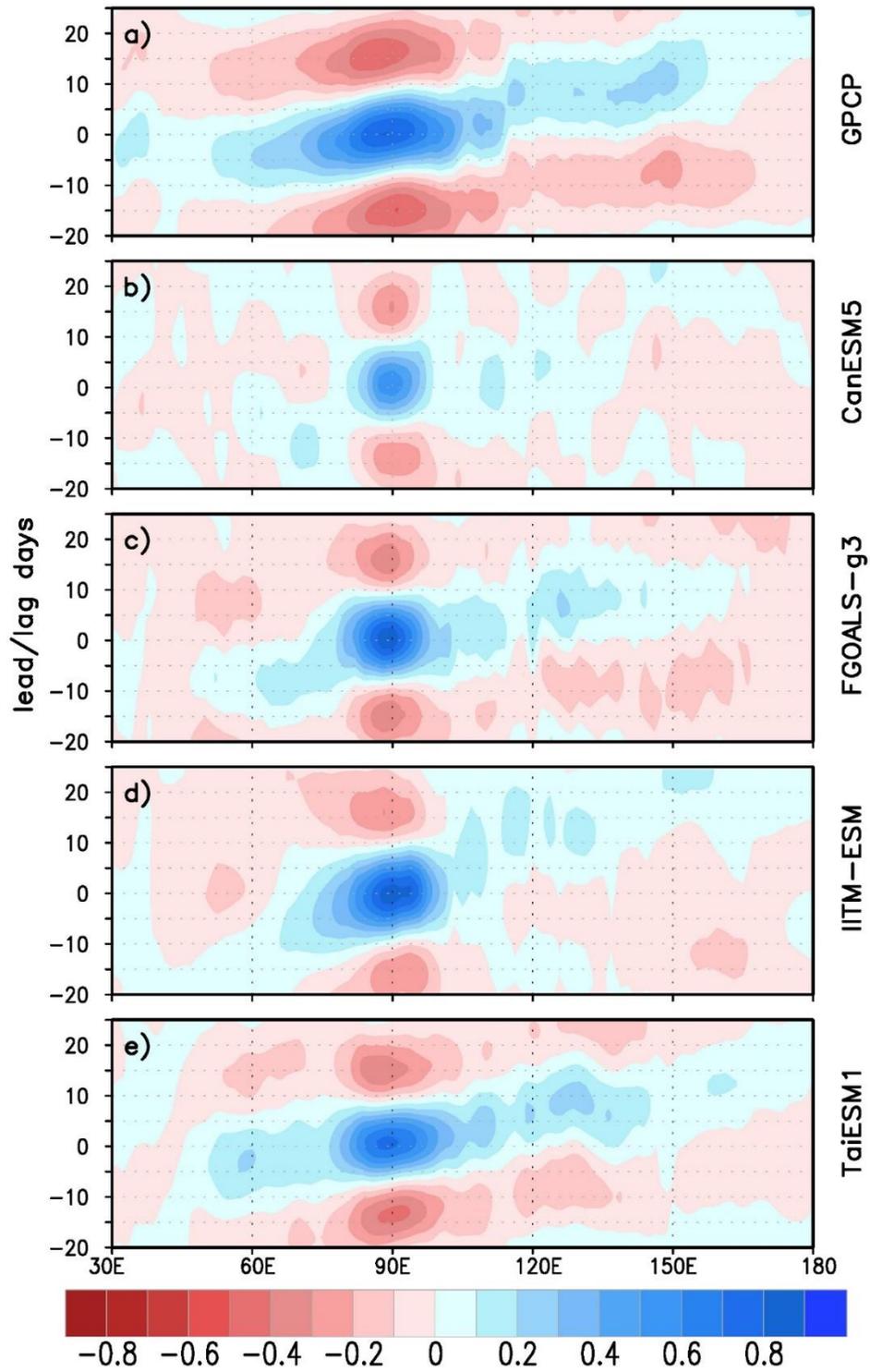

**Figure 14:** Same as Figure 13 but for boreal winter (Nov-Apr) season.

**Table.1:** Biases of rainfall (mm/day) over different parts of the south Asian monsoon region (SAMR). Positive(negative) value signifies wet (dry) bias.

| Region | CanESM5 | FGOALS-g3 | IITM-ESM | TaiESM1 |
|---|---|---|---|---|
| **Central India** | - 6.2 | - 5.4 | - 3.0 | - 1.3 |
| **Equatorial Indian Ocean** | +2.1 | +1.0 | -0.8 | -0.2 |
| **Bay of Bengal** | -2.9 | -1.8 | +0.7 | +0.9 |